# Spherical indentation on biological films with surface energy

Yue Ding, Wei-Ke Yuan and Gang-Feng Wang

Department of Engineering Mechanics, SVL, Xi'an Jiaotong University, Xi'an 710049, China

E-mail: wanggf@mail.xjtu.edu.cn

**Abstract**

Micro-/nano-indentations have been widely used to measure the mechanical properties of biological cells and tissues, but direct application of classical Hertzian contact model would lead to overestimation of elastic modulus due to the influence of finite thickness and surface energy. In this work, we analyze spherical indentation of biological films considering both large deformation and surface energy. The hyperelastic behavior of biological films is characterized by neo-Hookean model, and the influence of surface energy is addressed through finite element simulation. Based on dimensional analysis, the explicit expressions of load-depth relation accounting for film thickness, large deformation and surface energy are achieved for bonded or non-bonded films. Under a specific load, the consideration of large deformation increases the indent depth, while the finite thickness of films tends to decrease the indent depth, compared to the linear elastic Hertzian solution. More importantly, surface energy evidently alters the load-depth relation for micro-/nano-indentations, which reduces the indent depth and makes the films seemingly stiffer. These results provide a fundamental relationship to accurately extract the mechanical properties of biological films from indentation tests.

## 1. Introduction

Mechanical properties of biological cells and tissues are important factors in determining their cellular behaviours such as differentiation (1), aggregation, motility (2) and growth. Changes of cellular mechanical properties can be used as indicators for disease diagnosis (3, 4). Thus, it is quite meaningful to develop advanced methods to accurately evaluate these properties. Nowadays, micro-/nano-indentations through atomic force microscopy (AFM) have been extensively used to measure the mechanical properties of cells and tissues, such as elastic modulus, viscosity and hardness (4-7).

In experiments, to avoid errors caused by the uncertainty of contact point, the indent depth is required to be larger than a few hundred nanometers (8). Considering the fact that the thickness of cells and tissues is on the order of several microns, it is more proper to treat these biological samples as thin films instead of infinite elastic medium. In analyzing the load-depth curves of cells and tissues, the adoption of linear elastic Hertzian model has been proved to be inappropriate and would highly overestimate the elastic modulus (9). To extract the actual mechanical properties of films, Dimitriadis et al. derived the solutions for spherical indentations on finite thickness films bonded or non-bonded on substrate (9). For AFM conical indentation on thin samples, Gavara and Chadwick presented an analytical solution to exclude the bottom effect (10). However, in these works, the soft films are regarded as linear elastic solids and large deformation encountered in actual cases is neglected. Recently, Long et al. studied spherical indentation on films undergoing large deformation, but only bonded samples are analyzed (11).

For biological cells and soft materials, a variety of experiments have displayed that surface tension exerts a significant influence on their shape (12), cellular behavior (13), instability (14) and contact behavior (15). In recent years, surface effects on contact problems of linear elastic and hyperelastic materials have received extensive attentions. The indentation on an elastic half space has been studied by Hajji to measure the elasticity of inflated lobes (16). Based on the surface elasticity theory, He and Lim derived the surface Green function for an incompressible isotropic half space (17). Huang and Yu formulated the surface Green's functions for two dimensional problem (18). The surface Green function of the anti-plane case was given by Chen and Zhang (19). On the other hand, Wang and Feng derived the plane-strain elastic field induced by a concentrated force considering residual surface tension (20). Long and Wang investigated the spherical indentation on an elastic half space with surface tension (21). Gao *et al.* considered both the influence of surface elasticity and surface tension on spherical indentation (22). Xu et al. analyzed the contact between a hyperelastic half space and a rigid sphere considering both surface tension and adhesion (23). Recently, Ding et al. studied the spherical indentation on a hyperelastic half space (24) and spherical cells (25) with surface energy, respectively. However, surface effect on spherical indentation of soft films has not been investigated.

In this work, finite element simulations and dimensional analysis are adopted to investigate the spherical indentations on soft films with surface energy at large deformation. The results show that for biological films at large deformation, three correction factors including the large indent depth, film thickness and surface effect, should be introduced to modify the Hertzian solution. Simple explicit expression of load-depth relation is presented to extract the mechanical properties of biological films more accurately.

## 2. Finite element model

A biological film with thickness $h$ is indented through a rigid spherical indenter with radius $R$, as shown in Fig. 1. We consider two boundary conditions on the interface between the film and the rigid substrate: (*A*) the non-bonded (frictionless) condition with zero shear stress along the interface and (*B*) the bonded condition with zero displacement along the interface. In both cases, the normal displacement on the interface is prescribed to be zero. The load $P$ is applied on the top surface of thin film through the spherical indenter, which leads to an indent depth $d$ and a circular contact area with radius $a$. A cylindrical coordinate system is established here for convenience. The origin $O$ is located at the initial contact point and the $z$-axis is along the compressive direction. The finite element simulations are carried out by the commercial finite element method (FEM) software ABAQUS.

In our simulations, the biological film is characterized by the incompressible neo-Hookean model, which has been widely used to capture the hyperelastic properties of soft solids and biological materials (26, 27). The strain energy density $U$ of the neo-Hookean model is given by

$$U=\frac{E}{6}(I_1-3), \quad (1)$$

where $E$ is the elastic modulus and $I_1=\lambda_1^2+\lambda_2^2+\lambda_3^2$ is the first invariant of the principal stretches $\lambda_i$.

To introduce the influence of surface energy, we develop surface elements via the user subroutine UEL. In our calculations, a constant surface energy density $\gamma$ is assumed on the surface of the film, which corresponds to a constant surface tension. The Newton-Raphson method in ABAQUS is adopted here to seek the equilibrium state of this nonlinear contact problem. The details of FEM simulations with the incorporation of surface energy can be referred to our previous work (24).

The hyperelastic film is modelled as a circular layer with its radius being ten times of the thickness $h$, which is large enough to rule out the side boundary effect. The four-node bilinear axisymmetric quadrilateral hybrid reduced integration elements (CAX4RH) are used to discretize the film, and user-defined surface elements are attached on the top surface of the film, as shown in Fig. 2. On the symmetrical axis, the displacement along $r$-axis is restricted to be zero. On the bottom boundary, for the non-bonded film, the displacement normal to the interface $u_z$ and the shear stress $\sigma_{zr}$ are prescribed to be zero, and for the bonded film, all the displacements $u_z$ and $u_r$ are prescribed to be zero. Assume that the indenter is rigid and the contact between the indenter and the film is frictionless. The maximum indent depth is set as $d=0.4h$ and the radius of spherical indenter varies from $0.1h$ to $10h$. Here we consider three soft silicone films with elastic moduli $E = 85$ kPa, 250 kPa and 500 kPa, and the corresponding surface energy density $\gamma = 0.032$ N/m, 0.039 N/m and 0.035 N/m, respectively (15). Convergence tests have been carried out to ensure the accuracy of computational results.

## 3. Results and discussions

Firstly, we examine the influence of surface energy on the elastic field of films. When surface energy is taken into account for solids, the characteristic length that surface energy affects can be estimated by

$$s = \frac{2\gamma}{E^*}. \tag{2}$$

Here $E^*=E/(1-v^2)$ is the combined elastic modulus and $v$ the Poisson's ratio of the material. For the three considered soft materials, the intrinsic lengths are 0.564 μm, 0.234 μm and 0.105 μm, respectively. For contact problems, the relative significance of surface energy to bulk deformation can be indicated by the ratio of $s$ to contact radius $a$.

For an external load $P=0.1$ μN applied on a spherical indenter with radius $R=10$ μm, the normal stress $\sigma_z$ on the top surface of the non-bonded films are shown in Fig. 3. The normal stress from Hertzian prediction (28) is also plotted for comparison. For films without surface energy, the distribution of normal stress coincides with the linear elastic Hertzian contact model, even for hyperelastic solids at large deformation.

It can be found that the presence of surface energy smoothens the distribution of normal stress and induces nonzero value outside the contact region. Moreover, surface energy makes the bulk stress to vary smoothly across the contact edge.

Fig. 4 displays the normal displacement of non-bonded films around the contact region. For the cases $\gamma=0$, the normal displacement of non-bonded films is smaller than the Hertzian solution on the whole surface. For a specific load, the softer the film is, the larger the discrepancy of normalized normal displacement of thin film from the Hertzian prediction is, due to the finite thicknesses and large deformation. When surface energy is considered, the normal displacement inside the contact area will decrease furthermore, and thus the films appear to be stiffer. Similar variation trend of both normal stress and normal displacement can be observed for bonded films

Based on the comparison and analysis of the above results, we can appreciate that three factors, i.e. the thickness of film, the large deformation and the surface energy, contribute to the spherical indentation of biological films. Then we investigate the overall mechanical response of the spherical indentation on the hyperelastic films. For simplification, we firstly consider the influences of film thickness and large deformation with vanishing surface energy.

For a rigid sphere with radius $R$ indenting an elastic half space, Hertzian contact theory predicts the dependence of load $P_\text{H}$ on indent depth $d$ as

$$P_\text{H} = \frac{4}{3} E^* R^{0.5} d^{1.5}. \qquad (3)$$

For **a hyperelastic half space undergoing large deformation**, Zhang et al. modified the Hertzian solution as (29)

$$P_Z = \frac{4}{3} E^* R^{0.5} d^{1.5} \left(1 - 0.15 \frac{d}{R}\right). \tag{4}$$

This result implies that large deformation tends to increase the indent depth for a given load, depending on the ratio $d/R$.

For spherical indentation on **a non-bonded incompressible linear elastic film**, Dimitriadis et al. provided the relation between load and indent depth as (9)

$$P_D^s = \frac{4}{3} E^* R^{0.5} d^{1.5} \left(1 + 0.884\chi + 0.781\chi^2 + 0.386\chi^3 + 0.0048\chi^4\right), \tag{5}$$

in which $\chi = \sqrt{Rd}/h$ indicates the influence of film thickness. This solution is applicable for $\chi \leq 1$. It reveals that the indent depth will decline as the film thickness decreases under a specific load.

Considering the above solutions Eqs. (3)-(5) and adopting dimensional analysis, the load $P_{nH}^s$ of spherical indentation on a hyperelastic film at large deformation can be expressed as

$$P_{nH}^s = \frac{4}{3} E^* R^{0.5} d^{1.5} f_{nH}^s \left(\frac{d}{R}, \chi\right), \tag{6}$$

where $f_{nH}^s$ is a function depending on the dimensionless parameters $d/R$ and $\chi$.

Numerous finite element simulations with different indenter radii, film thicknesses and material parameters are carried out to achieve the explicit load-depth relation of spherical indentation on hyperelastic films. To display the influence of film thickness, the variation of load $P_{nH}^s$ normalized by Zhang's solution $P_Z$ is plotted with respect to $\chi$ in Fig. 5. It is found the total external load $P_{nH}^s$ can be well approximated by

$$P_{nH}^s = \frac{4}{3} E^* R^{0.5} d^{1.5} \left(1 - 0.15 \frac{d}{R}\right) \left[\left(1 + 0.97\chi + 0.91\chi^2\right)\left(1 - 0.15\chi\right)\right]. \tag{7}$$

**This expression takes into account both the large deformation and the film thickness**. For comparison, the linear elastic solution of film, $P_D^s$, given by Dimitriadis et al., is also shown, which is normalized by Hertzian solution $P_H$ to demonstrate the thickness effect. For $\chi<0.5$, Eq. 7 coincides well with Dimitriadis et al's result (Eq. 5) for linear elastic solid. However, as $\chi$ increases, the discrepancy between our model (Eq. 7) and Dimitriadis et al's solution (Eq. 5) becomes evident due to large deformation and a lower load is predicted by Eq. 7 for a given indent depth.

For **incompressible elastic films bonded on substrate**, Dimitriadis et al. gave the load-depth relation at small deformation as

$$P_D^b = \frac{4}{3} E^* R^{0.5} d^{1.5} \left(1 + 1.133\chi + 1.283\chi^2 + 0.769\chi^3 + 0.0975\chi^4\right). \quad (8)$$

For **hyperelastic films at large deformation**, Long et al. (11) modified the load-depth relation as

$$P_L^b = \frac{4}{3} E^* R^{0.5} d^{1.5} \frac{1 + 2.3\chi^3}{1 + 1.15\chi + \alpha\chi^3 + \beta\chi^6}. \quad (9)$$

where $\alpha = 10.05 - 0.63\sqrt{h/R}\left(3.1 + h^2/R^2\right)$ and $\beta = 4.8 - 4.23 h^2/R^2$.

We also calculate the load-depth curve for spherical indentation on bonded films, as shown in Fig. 5. A new fitting relation is advanced as

$$P_{nH}^b = \frac{4}{3} E^* R^{0.5} d^{1.5} \left(1 - 0.15\frac{d}{R}\right)\left[\left(1 + 0.6\chi + 2.5\chi^2 + 0.45\chi^3\right)\left(1 + 0.07\chi\right)\right] \quad (10)$$

It is seen that the new proposed relation Eq. 10 can capture the FEM results, similar as Long et al's solution Eq. 9, while in a simpler expression. For large indentation as $\chi>0.5$, the load curve is higher than that of linear elastic case, in contrast to the case for non-bonded films. As $\chi$ increases, corresponding to the increase of indent depth or the decrease of film thickness, the difference induced by boundary conditions gets larger. For example, the load for bonded films can be three times larger than that for non-bonded films when $\chi=2$.

When surface energy is considered, the mechanical response of indentation is altered and the surface effect can be indicated by the ratio of the intrinsic length $s$ to Hertzian contact radius $\sqrt{Rd}$ as demonstrated in previous investigation (24). According to dimensional analysis, for spherical indentation on a soft film with surface energy, the external load $P_{\text{nHs}}^{s}$ for non-bonded films can be expressed as

$$P_{\text{nHs}}^{s} = P_{\text{nH}}^{s} f_{\text{nHs}}^{s}\left(\frac{\sqrt{Rd}}{s}, \chi\right), \tag{11}$$

where $P_{\text{nH}}^{s}$ is the load for the case without surface energy given by Eq. 7, and $f_{\text{nHs}}^{s}$ is a dimensionless function of parameters $\chi$ and $\sqrt{Rd}/s$.

A variety of cases for different indenter radii and hyperelastic films with different values of thickness, elastic modulus and surface energy are simulated to achieve the load-depth relation. Fig. 6 displays the FEM results of normalized load with respect to $\sqrt{Rd}/s$. It is found that the dependence of load on indent depth can be well fitted by

$$P_{\text{nHs}}^{s} = P_{\text{nH}}^{s}\left\{1 + 0.88\left(\frac{\sqrt{Rd}}{s}\right)^{-0.87}(1 - 0.25\chi)\right\} \tag{12}$$

When $\sqrt{Rd}$ is much larger than the intrinsic length $s$, the influence of surface energy is negligible and the load can be predicted by Eq. 7. When $\sqrt{Rd}$ is comparable or smaller than the intrinsic length $s$, which meets the cases for

micro-/nano-indentations on biological tissues or soft solids, surface energy will evidently affect the relation between load and indent depth. To yield a given indent depth, the existence of surface energy predicts that a higher load is required, implying a stiffening effect of surface energy. As the ratio of $\sqrt{Rd}$ to $s$ decreases, the surface effect on the applied load becomes more evident. The overlook of surface energy will lead to overestimating the elastic modulus of biological samples.

Based on dimensional analysis mentioned above, finite element simulations for bonded films have also been carried out as shown in Fig. 7. The FEM results can be reproduced well by the function

$$P_{\text{nHs}}^b = P_{\text{nH}}^b \left\{ 1 + 0.88 \left( \frac{\sqrt{Rd}}{s} \right)^{-0.87} (1 - 0.4\chi) \right\} \tag{13}$$

**The effects of both large deformation and surface energy are included** in Eqs. 12 and 13, which provides a more accurate relation to characterize the compressive response of hyperelastic films under large deformation. As the film becomes thicker, the influence of thickness on load-depth relation diminishes. When the thickness of the film is much larger than the radius of indenter (i.e. $\chi \ll 1$), the load-depth relation (both Eqs. 12 and 13) reduces to the solution of spherical indentation on hyperelastic half space as we investigated previously (24).

## 4. Conclusions

The spherical indentation on hyperelastic films with surface energy at large deformation has been investigated. Through finite element method and dimensional analysis, explicit expressions of load–depth relation for both non-bonded and bonded films have been achieved. Comparing with linear elastic predictions, our models considering the finite thickness and large deformation predict a smaller load for non-bonded films, but a larger load for bonded films. When the contact radius is close to the intrinsic length *s*, the effect of surface energy should be considered, which tends to resist the deformation for both non-bonded and bonded films. For the elastic field within the contact region, both the normal displacement and normal stress on the top surface of film decrease due to the existence of surface energy. Our results provide a more accurate expression to measure the mechanical properties of films and are helpful to understand the mechanical response for soft films under indentation.

**ACKNOWLEDGEMENTS**

Supports from the National Natural Science Foundation of China (Grant No. 11525209) are acknowledged.

**Figure captions:**

**FIGURE 1.** Schematic of spherical indentation on hyperelastic film of thickness *h* with (*A*) non-bonded condition and (*B*) bonded condition. The load *P* is applied through the spherical tip with radius *R* and leads to an indent depth *d* and a circular contact area with radius *a*.

**FIGURE 2.** Finite element model of the spherical indentation on film. The zoom-in view of mesh near the contact region is also presented.

**FIGURE 3.** Normal stress distribution around the contact area of non-bonded films. The lines represent the films with surface energy, and symbols refer to the cases without surface energy. Hertzian solution is also plotted here for comparison.

**FIGURE 4.** Normal displacement on the surface of non-bonded films. The dashed lines show the cases without surface energy, and lines with symbols represent the films with surface energy.

**FIGURE 5.** Relation between load and depth for film at large deformation without surfacen energy. The circular symbols are FEM results, which can be fitted well by solid lines (Eq. 7 and 10). The dashed lines are predictions from Dimitriadis et al. and triangular symbol represent the solution from Long et al.

**FIGURE 6.** Load-depth relation for non-bonded film with surface energy. The solid line is predicted by Eq. 12 and agrees well with the symbols produced by FEM.

**FIGURE 7.** Load-depth relation for bonded film with surface energy. The symbols are FEM results and the solid line represent the prediction from Eq. 13.

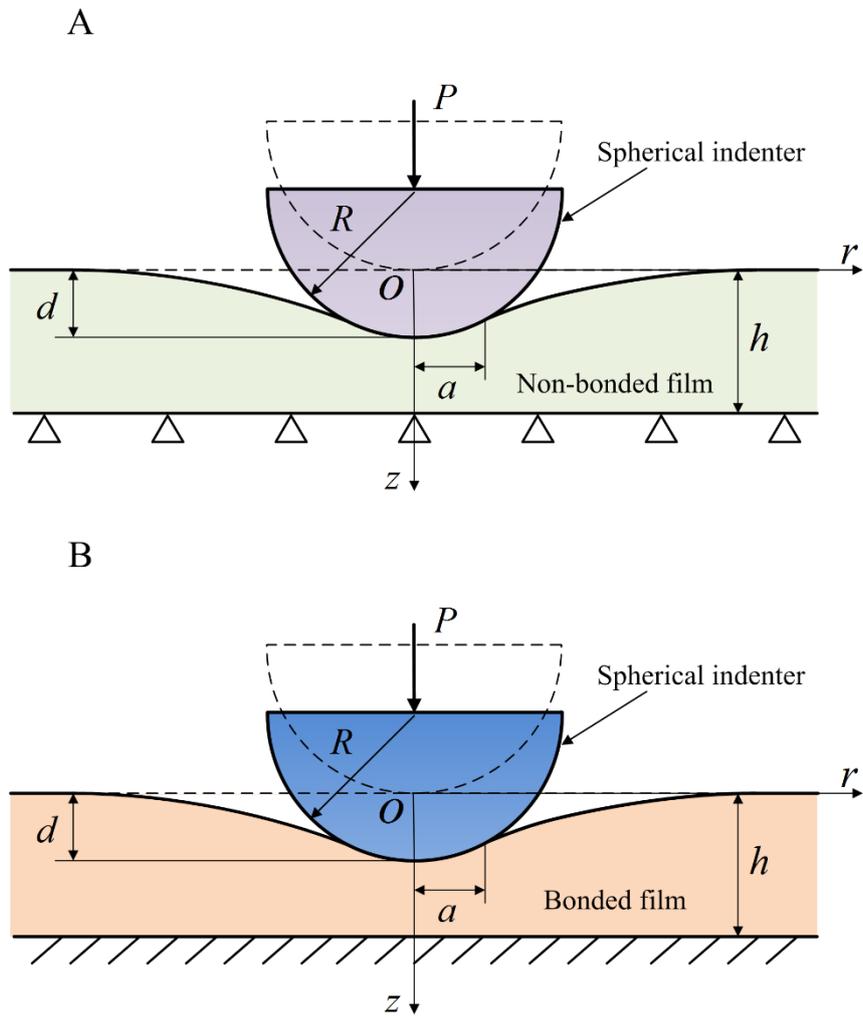

FIGURE 1

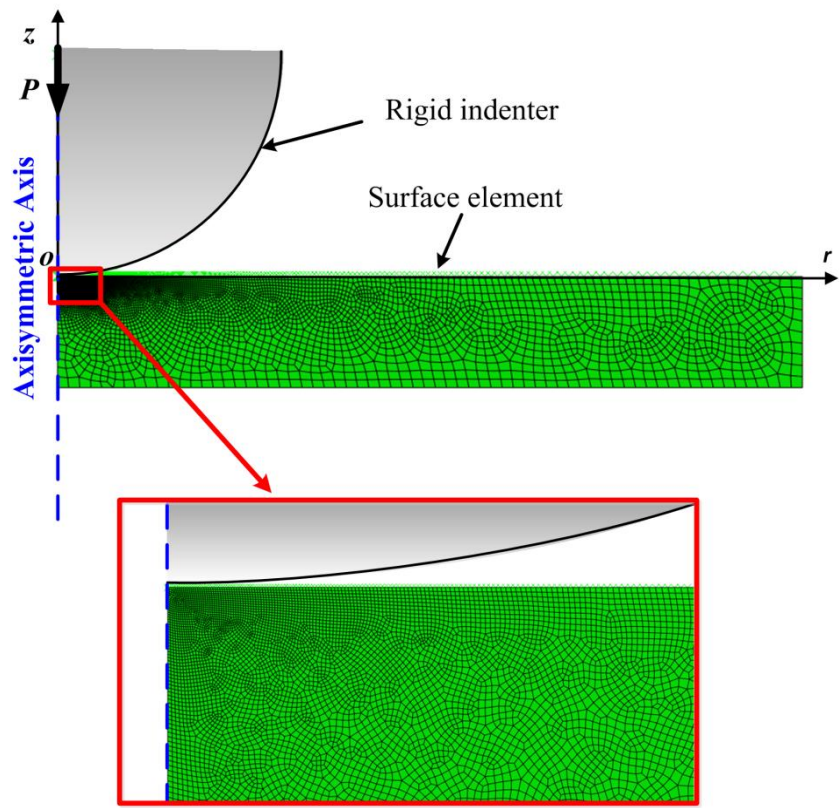

FIGURE 2

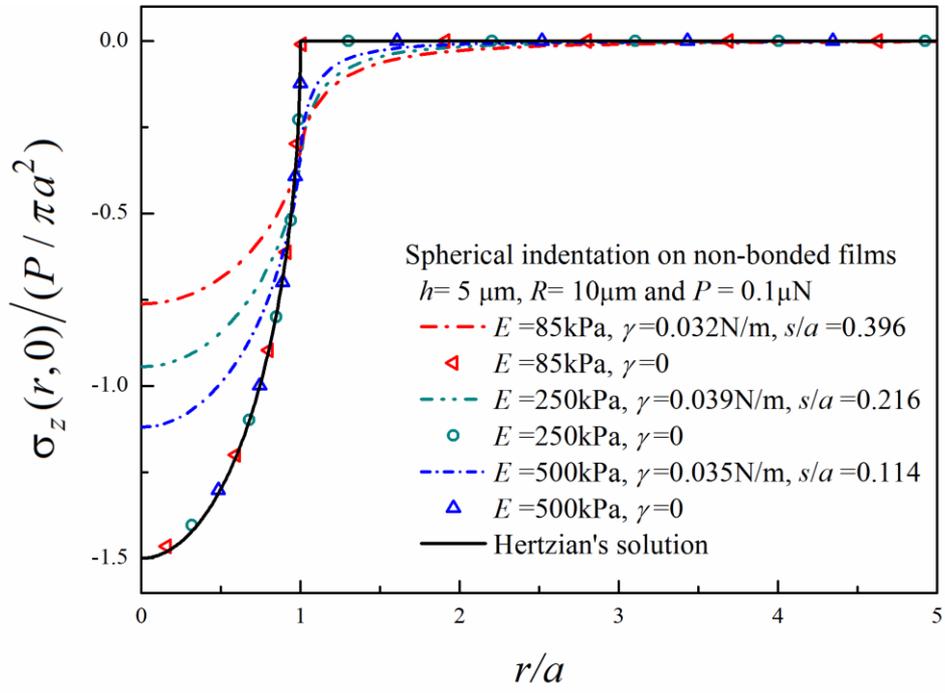

FIGURE 3

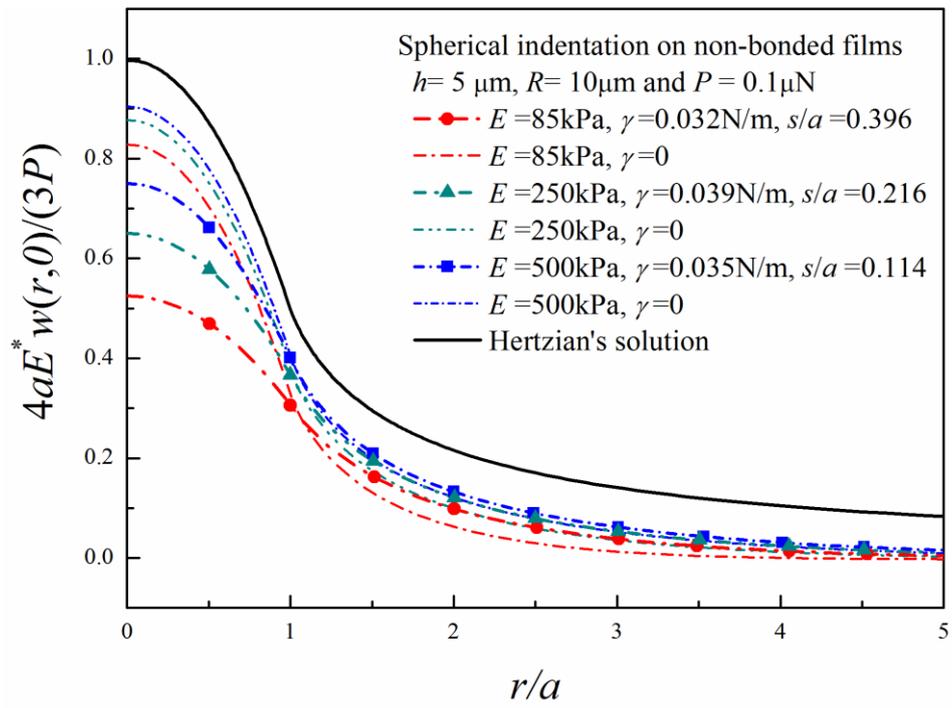

FIGURE 4.

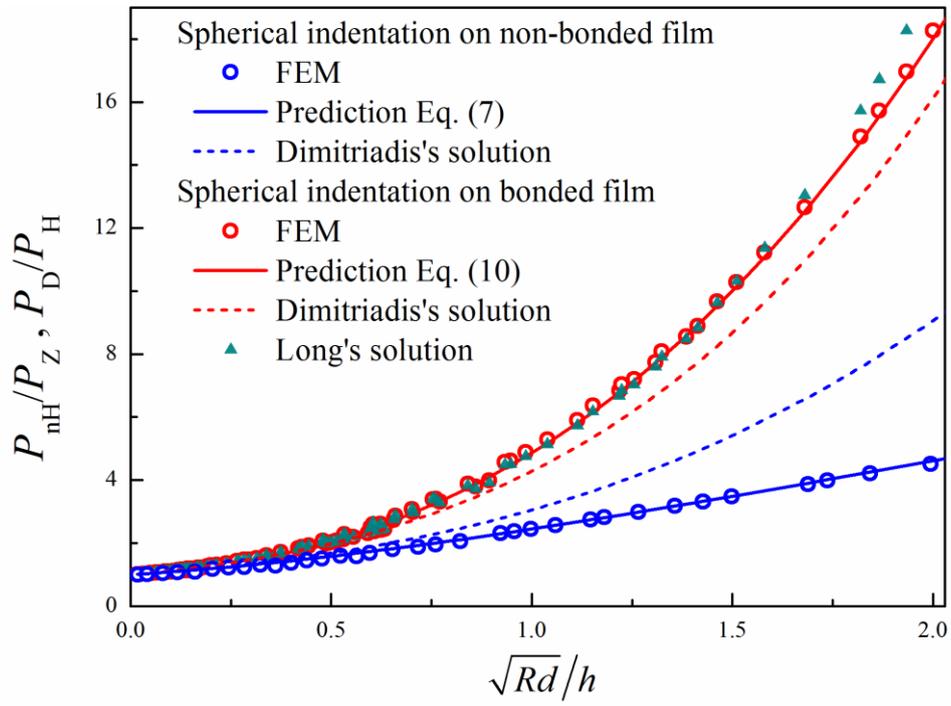

FIGURE 5

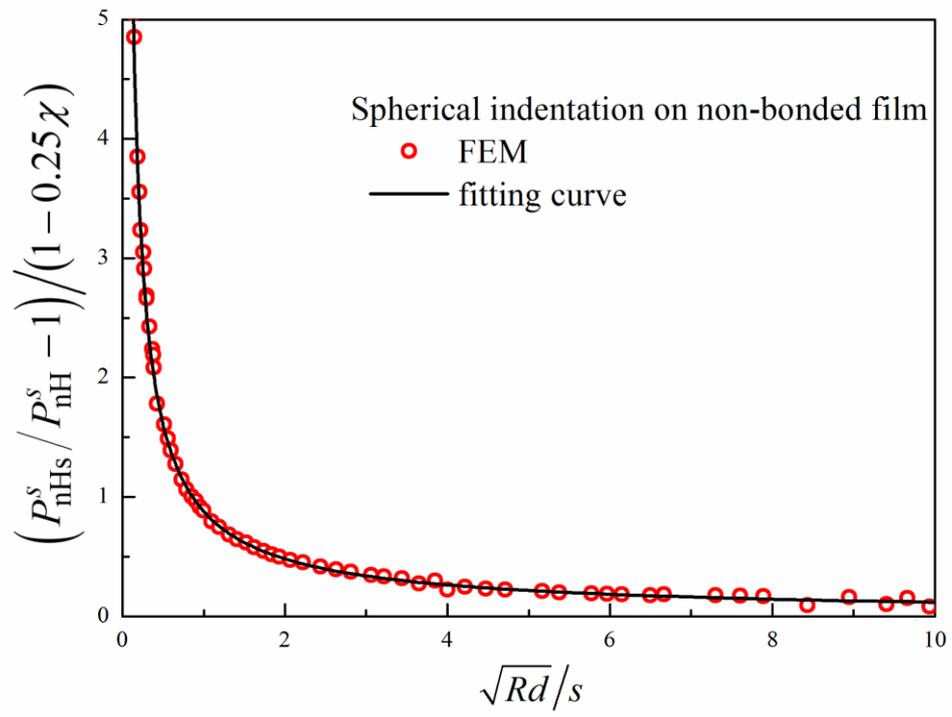

FIGURE 6

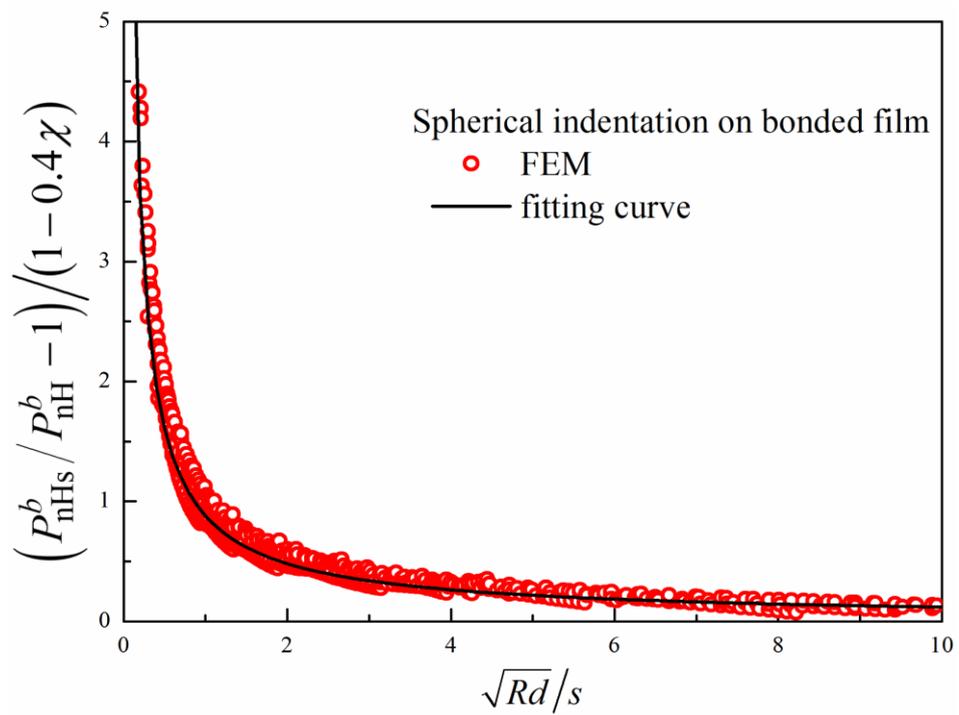

FIGURE 7